\documentclass[journal,letterpaper]{IEEEtran}
\usepackage{amsmath}
\usepackage{amssymb}
\usepackage{amsfonts}
\usepackage{graphicx}
\usepackage{epsfig}
\usepackage{subfigure}
\usepackage{psfrag}
\usepackage{cite}
\usepackage{color}
\usepackage[letterpaper,margin=0.8in]{geometry}
\usepackage{multicol}

\usepackage{dblfloatfix}

\begin{document}
\title{Massive Connectivity with Massive MIMO--Part II:
Achievable Rate Characterization} 
\author{Liang Liu, \IEEEmembership{Member,~IEEE} and Wei Yu, \IEEEmembership{Fellow,~IEEE}

\thanks{Manuscript received June 17, 2017, revised November 9, 2017 and January 30, 2018, accepted March 8, 2018. The materials in this paper have been
presented in part at the IEEE International Symposium on Information Theory
(ISIT), Aachen, Germany, July 2017 \cite{LiangISIT}. The associate editor coordinating the review of this paper and approving it for
publication was Dr. Mathini Sellathurai. This work is supported by Natural Sciences and Engineering
Research Council (NSERC) of Canada via a discovery grant, a Steacie Memorial Fellowship
and the Canada Research Chairs program.}
\thanks{The authors are with The Edward S. Rogers
Sr. Department of Electrical and Computer Engineering, University of Toronto,
Toronto, Ontario, Canada, M5S3G4,
(e-mails:lianguot.liu@utoronto.ca; weiyu@comm.utoronto.ca).}}

\maketitle

\begin{abstract}
This two-part paper aims to quantify the cost of device activity
detection in an uplink massive connectivity scenario with a large
number of devices but device activities are sporadic.  Part I of
this paper shows that in an asymptotic massive multiple-input
multiple-output (MIMO) regime, device activity detection can always be
made perfect. Part II of this paper subsequently shows that despite
the perfect device activity detection, there is nevertheless
significant cost due to device detection in terms of overall achievable
rate, because of the fact that non-orthogonal pilot sequences have to
be used in order to accommodate the large number of potential devices,
resulting in significantly larger channel estimation error as compared
to conventional massive MIMO systems with orthogonal pilots.
Specifically, this paper characterizes each active user's achievable
rate using random matrix theory under either maximal-ratio combining (MRC) or minimum mean-squared
error (MMSE) receive beamforming at the base-station (BS), assuming the
statistics of their estimated channels as derived in Part I.
The characterization of user rate also allows the optimization
of pilot sequences length.
Moreover, in contrast to the conventional massive MIMO system,
the MMSE beamforming is shown to achieve much higher rate
than the MRC beamforming for the massive connectivity
scenario under consideration.  Finally, this paper illustrates
the necessity of user scheduling for rate maximization when the number
of active users is larger than the number of antennas at the BS.

\end{abstract}

\begin{IEEEkeywords}
Beamforming, massive connectivity, massive multiple-input multiple-output (MIMO),
random matrix theory, large-system analysis,
Internet-of-Things (IoT), machine-type communications (MTC).
\end{IEEEkeywords}

\IEEEpeerreviewmaketitle

\newtheorem{corollary}{Corollary}
\newtheorem{definition}{Definition}
\newtheorem{lemma}{Lemma}
\newtheorem{theorem}{Theorem}
\newtheorem{proposition}{Proposition}
\newtheorem{remark}{Remark}
\newcommand{\mv}[1]{\mbox{\boldmath{$ #1 $}}}

\section{Introduction}\label{sec:Introduction}
\subsection{Motivation}

Motivated by the emerging Internet-of-Things (IoT) and machine-type
communications (MTC) applications, this two-part paper studies the
uplink communication in a massive multiple-input multiple-output (MIMO) single-cell system, in which a base-station (BS) is equipped with a large number of antennas to serve a massive number of
devices with sporadic traffic. Specifically, the BS
is equipped with $M$ antennas, serving $N$ potential devices, out of
which $K$ are active at any given time. A two-phase multiple-access
scheme is adopted in which within each coherence time of length $T$,
the active users send their pilot sequences during the first $L<T$
symbols for device activity detection and channel estimation in the
first phase, while send their data
messages during the remaining $T-L$ symbols in the second phase.

A key challenge of the above system is that due to the limited
coherence time, only non-orthogonal pilot sequences can be assigned to
the users, as typically $N \gg L$. The main objective of Part I of
this paper \cite{LiuPart1} is to quantify the performance of device activity
detection and channel estimation when \emph{randomly generated non-orthogonal
pilot sequences} are assigned for each device. Part II of this paper examines
its impact on the overall achievable data rate for this massive connectivity
system with massive MIMO.

Part I of this paper \cite{LiuPart1} shows that the user activity
detection and channel estimation problem in the first phase can be cast
as a compressed sensing problem that takes advantage of the sparsity
in device activity, for which the approximate message passing (AMP)
algorithm \cite{donoho_amp,baron_MMV,bayati_montanari_amp,rangan3} can
be used to solve the above problem. Specifically, Part I of
this paper designs a minimum mean-squared error (MMSE) denoiser in
a vector form of the AMP algorithm for user activity detection and
channel estimation based on the statistics of the channel, and shows
that in certain asymptotic regime where $K,N,L$ all go to infinity,
the probabilities of missed detection and false alarm as well as the
statistical distributions of the active users' estimated channels can
be characterized analytically. 
Interestingly, it is shown that the MMSE-based AMP algorithm is
capable of driving the user detection error probability down to zero
as the number of BS antennas $M$ goes to infinity.
Thus, massive MIMO is naturally suited for massive connectivity.

%

Part II of the paper leverages the above perfect user activity
detection result as well as the statistical distributions of the
estimated channels to characterize in closed-form the overall achievable
rates under the aforementioned two-phase transmission protocol with
either the maximal-ratio combining (MRC) or the MMSE beamforming at
the BS, again in the massive MIMO regime as $M$ goes to infinity.
Our main conclusion is that despite perfect detection, there is
nevertheless significant cost on user achievable rate due to massive device detection because
the use of non-orthogonal pilot sequences results in larger channel estimation errors.



\subsection{Prior Work}

Massive MIMO systems \cite{marzetta}, where each BS is equipped with
a large number (sometimes in the order of 100's) of antennas, have emerged as a key technology
for achieving dramatic spectral efficiency gains in future wireless
systems. In a single-cell system where the number of antennas at the
BS is much larger than that of the users, the channels of different
users become asymptotically orthogonal under the so-called
``favorable'' propagation conditions \cite{marzetta}. As a result,
simple matched filter (MF) processing, such as maximal-ratio
transmission (MRT) in the downlink and MRC in the uplink, is already
asymptotically optimal for maximizing the user rate, assuming perfect channel state
information (CSI) \cite{marzetta}.  Moreover, it is shown in
\cite{larsson} that each single-antenna user in a massive MIMO system
can scale down its transmit power proportional to the number of
antennas at the BS to get the same rate performance
as a corresponding single-input single-output (SISO) system, assuming perfect CSI.

Despite its promises, massive MIMO system is also faced with many
practical challenges, chief among which is channel estimation
\cite{larsson14}. Channel training for the uplink MIMO system should
typically be done with orthogonal pilot sequences within each cell;
further the optimal training length in time should be the same as the
number of transmit antennas in uncorrelated Rayleigh fading channels
\cite{hassibi_hochwald}.  With this channel training strategy, the
user rates achieved by the MRC beamforming and MMSE beamforming at the
BS are characterized in \cite{larsson,debbah} utilizing the random
matrix theory, where it is shown that even with imperfect CSI, the
throughput achieved by the MRC beamforming is very close to that of
MMSE beamforming in the uplink massive MIMO system. We remark that
channel training is even more challenging in the downlink massive MIMO
system, especially when the system operates in the frequency-division duplex
(FDD) mode where channel reciprocity does not hold between uplink and
downlink.  
Many sophisticated schemes have been proposed for this long-standing
problem in the downlink FDD massive MIMO system \cite{love14,larsson14}.
Finally, we mention
that in a multi-cell system, the non-orthogonality of the pilot
sequences in nearby cells causes pilot contamination, which then
becomes the dominant impairment in the asymptotic massive MIMO regime
\cite{marzetta}.

In contrast to the conventional massive MIMO literature, this paper
points out that channel training can be a limiting factor even in the
single-cell uplink scenario, when massive number of devices are
involved.  This is because when the total number of devices is much
larger than the number of BS antennas, it is impossible to assign
orthogonal pilot sequences to each device. Part I of this paper \cite{LiuPart1} deals
with device activity detection.
In this Part II of the
paper, we aim to quantify the cost of non-orthogonal pilots for
channel estimation and subsequently the overall achievable rate.
One of the consequences of our result is that MMSE beamforming is necessary for
maximizing the user rate, because of the fact 
that the inter-user interference cannot be effectively canceled by a
simple MRC operation when the number of active users is comparable to the number of antennas at the BS.

\subsection{Main Contributions}

This two-part paper provides an analytical performance
characterization of the two-phase transmission protocol in a
single-cell massive connectivity scenario with massive MIMO, in which
the active users send their non-orthogonal pilot sequences to the BS
simultaneously for user activity detection and channel estimation in
the first phase, then transmit data to the BS for information decoding
in the second phase, within the same coherence time. The main
contributions of Part II of this paper are as follows.

Based on the user activity detection and channel estimation statistics
results of Part I of this paper and also based on techniques from
random matrix theory, we characterize the user achievable rate for
both the cases of MRC and MMSE beamforming at the BS, in an asymptotic
limit where the number of antennas at the BS and the number of users
both go to infinity, while keeping their ratio fixed.
By comparing to the case with prior information of user activity at
the BS, it is shown that despite the guaranteed success in activity
detection, the non-orthogonality of pilot sequences can nevertheless
lead to significantly larger channel estimation error as compared to
the conventional massive MIMO system, thus limiting the overall
achievable transmission rate. We quantify this cost and illustrate
that the optimal pilot sequence length in a massive connectivity
system should be longer than that in conventional massive MIMO system
for maximizing the overall transmission rate.

This paper shows that the massive connectivity system also possesses
other fundamental differences as compared to the conventional massive
MIMO system with a small number of users. First, the user rate is
finite due to inter-user interference, even in a single-cell massive
MIMO system with infinite number of antennas and without pilot
contamination from other cells. Second, the user rate achieved by the
MMSE beamforming at the BS is significantly higher than that achieved
by the MRC beamforming. At last, we show that in an overloaded system
where the number of active users is much larger than that of the antennas at
the BS, user scheduling can significantly improve the overall
transmission rate if the MMSE beamforming is applied at the BS.

\subsection{Organization}
The rest of Part II of this paper is organized as follows.
Section \ref{sec:System Model} describes the system model for massive
connectivity and introduces the two-phase transmission protocol for
user detection, channel estimation, and data transmission.
Section \ref{sec:AMP for User Activity Detection and Channel Estimation in Massive MIMO}
reviews the vector AMP algorithm and its performance in terms of user activity
detection and channel estimation derived in Part I of this paper;
Section \ref{sec:Phase II: Data Transmission} analyzes user achievable rate with the MRC and
MMSE beamforming at the BS with or without user scheduling;
Section \ref{sec:The Cost of User Detection on Users' Rates} investigates the cost of user
activity detection on user rate; Sections \ref{sec:The Optimal Length of Pilot Sequences} and \ref{sec:Optimization of User Scheduling}
optimize the pilot sequence length and number of scheduled intervals to maximize the user sum rate, respectively;
Section \ref{sec:Numerical Examples} provides the numerical
simulation results pertaining to user achievable rate. Finally, Section \ref{sec:Conclusion} concludes
the paper and points out several future directions.


\subsection{Notation}

Scalars are denoted by lower-case letters, vectors by bold-face
lower-case letters, and matrices by bold-face upper-case letters.
The identity matrix and the all-zero matrix of appropriate
dimensions are denoted as $\mv{I}$ and $\mv{0}$, respectively.
For a matrix $\mv{M}$ of arbitrary size, $\mv{M}^{H}$ and $\mv{M}^{T}$
denote its conjugate transpose and transpose, respectively.
The expectation operator is denoted as $\mathbb{E}[\cdot]$.
The distribution of a circularly symmetric complex Gaussian (CSCG)
random vector with mean $\mv{x}$ and covariance matrix $\mv{\Sigma}$
is denoted by $\mathcal{CN}(\mv{x},\mv{\Sigma})$; 
the space of complex matrices of size $m\times n$ is denoted as
$\mathbb{C}^{m \times n}$.

\section{System Model}\label{sec:System Model}

The overall system model is as introduced in Part I of this two-part paper \cite{LiuPart1}.
The channel input-output relationship for the uplink
communication in a single cell consisting of $N$ single-antenna users and one BS with $M$ antennas is given as:
\begin{align}\label{eqn:channel}
\mv{y}= \sum_n \mv{h}_n \alpha_n x_n + \mv{z} =
\sum\limits_{k\in \mathcal{K}}\mv{h}_k x_k+\mv{z},
\end{align}
where $x_n \in \mathbb{C}$ with a power $\mathbb{E}|x_n|^2 = \rho$ is the transmit signal of user $n$,
$\mv{h}_n\in \mathbb{C}^{M\times 1}\sim \mathcal{CN}(\mv{0},\beta_n\mv{I})$ denotes the complex
uplink channel vector from user $n$ to the BS with a path-loss exponent $\beta_n$ known by the BS, $\mv{z}
\in \mathbb{C}^{M\times 1} \sim \mathcal{CN}(\mv{0}, \sigma^2 \mv{I})$ is the additive white
Gaussian noise (AWGN) vector at the BS, and $\mv{y}\in
\mathbb{C}^{M\times 1}$ is the received signal. Here in (\ref{eqn:channel}),
$\alpha_n$'s are the user activity indicators used to model the sporadic traffic pattern of massive connectivity, i.e., $\alpha_n=1$ if user
$n$ is active at one coherence time, and $\alpha_n=0$ otherwise, $n=1,\cdots,N$. At last, $\mathcal{K}$ is the set of active users within a coherence
block, i.e., $\mathcal{K}=\{n:\alpha_n=1, n=1,\cdots,N\}$, with a cardinality $K=|\mathcal{K}|$.

Within each coherence time with $T$ symbols, we adopt the following
two-phase multiple access scheme: in the first phase of length $L$
symbols, the BS conducts user activity detection and channel
estimation based on the pilot sequences from the active users; in the
second phase, the BS decodes user messages based on the estimated
channels in the previous phase.  The transmitted signals of the active
users are assumed to be synchronized in both phases. The key point
here is that in a massive connectivity system with $N>L$, it is
impossible to assign orthogonal pilots to all the potential users.
In this paper, we assume a non-orthogonal pilot sequence assignment
strategy in which each user $n$ is allocated to a pilot $\mv{a}_n\in
\mathbb{C}^{L\times 1}$ whose entries are generated from independently and identically distributed (i.i.d.)
complex Gaussian distribution with zero mean and variance $1/L$.


\section{User Activity Detection and Channel Estimation in Massive
MIMO Regime}
\label{sec:AMP for User Activity Detection and Channel Estimation in Massive MIMO}

The AMP algorithm is effective for device activity detection and
channel estimation for the massive connectivity scenario. This section
first summarizes the main analysis in \cite{LiuPart1}, then
further derives an analytic expression for channel estimation error
for system parameter regime of most interest, which is useful for subsequent
characterization of the cost of non-orthogonal pilot sequences on user
rate and for optimization of the pilot sequence length for rate maximization.

\subsection{AMP for Activity Detection and Channel Estimation}

Consider the first phase of massive device transmission in which each
user sends its pilot sequence synchronously through the channel. Define
$\rho^{\rm pilot}$ as the identical transmit power of the active users
in the first transmission phase. The transmit signal of user $n$ can be
expressed as $\alpha_n \sqrt{\xi } \mv{a}_n$, where $\xi  = L \rho^{\rm pilot} $
denotes the total transmit energy of each active user in the first phase. The received signal at the BS is then
\begin{equation}
\label{eqn:channel_training}
\mv{Y} =\sqrt{\xi }\mv{A}\mv{X}+\mv{Z},
\end{equation}
where $\mv{Y} \in \mathbb{C}^{L\times M}$ is the matrix of
received signals across $M$ antennas over $L$ symbols, $\mv{A}=[\mv{a}_1, \cdots, \mv{a}_N]$
is the collection of user pilot sequences,
$\mv{X}=[\mv{x}_1, \cdots, \mv{x}_N]^T$ is the collection of user
equivalent channels $\mv{x}_n=\alpha_n\mv{h}_n$'s, and
$\mv{Z}=[\mv{z}_1,\cdots,\mv{z}_M]$ with
$\mv{z}_m \sim \mathcal{CN}(\mv{0},\sigma^2\mv{I})$, $\forall m$,
is the independent AWGN at the BS.
As $\mv{X}$ is row sparse, Part I of this paper proposes to use the
MMSE-based vector AMP algorithm to recover $\mv{X}$ based on the noisy
observation $\mv{Y}$. 
More details on the implementation of the vector AMP algorithm can
be found in \cite{LiuPart1}.

The main result of \cite{LiuPart1} is an analytical characterization
of the user activity detection and channel estimation performance
using the vector AMP algorithm in the asymptotic regime where
$L,K,N\rightarrow \infty$, while their ratios converge to some fixed
positive values $N/L \rightarrow \omega$ and $K/N \rightarrow
\epsilon$ with $\omega, \epsilon \in (0,\infty)$, while keeping
the total transmit power fixed at $\xi $. Specifically, for user activity detection, we show that in the above asymptotic regime,
the probabilities of missed detection (a user is active but is declared as inactive) and false alarm (a
user is inactive but is declared as active) by the MMSE-based AMP algorithm both converge to zero exponentially as the number of antennas
at the BS, i.e., $M$, goes to infinity.

Moreover, for channel estimation, after the convergence of
the vector AMP algorithm,
the covariance matrices of the estimated channel of an active user
$k\in \mathcal{K}$, denoted by $\hat{\mv{h}}_k$, and the
corresponding channel estimation error, denoted by
$\Delta \mv{h}_k=\mv{h}_k-\hat{\mv{h}}_k$, are given, respectively, by
\begin{align}
& {\rm Cov}(\hat{\mv{h}}_k,\hat{\mv{h}}_k)=\upsilon_k(M)\mv{I}, \label{eqn:channel_estimate 1} \\
& {\rm Cov}(\Delta \mv{h}_k,\Delta \mv{h}_k)=\Delta \upsilon_k(M)\mv{I}, \label{eqn:channel_error 1}
\end{align}where $\upsilon_k(M)$ and $\Delta \upsilon_k(M)$ respectively converge to as the number of antennas at the BS goes to infinity:
\begin{align}
& \lim\limits_{M\rightarrow \infty}\upsilon_k(M)=\frac{\beta_k^2}{\beta_k+\tau_\infty^2}, \label{eqn:channel_estimate 2} \\
& \lim\limits_{M\rightarrow \infty}\Delta \upsilon_k(M)=\frac{\beta_k\tau_\infty^2}{\beta_k+\tau_\infty^2}. \label{eqn:channel_error 2}
\end{align}
In (\ref{eqn:channel_estimate 2}) and (\ref{eqn:channel_error 2}), $\tau_\infty^2$ is the fixed-point solution to the following
simplified state evolution of the AMP algorithm as $M\rightarrow\infty$:
\begin{align}
\label{eqn:state evolution scalar form 0}
& \tau_0^2=\frac{\sigma^2}{\xi}+\omega \epsilon \mathbb{E}_\beta[\beta], \\
& \tau_{t+1}^2=\frac{\sigma^2}{\xi } +
\omega \epsilon \mathbb{E}_{\beta}\left[\frac{\beta\tau_t^2}{\beta+\tau_t^2}\right], ~~~ t\geq 0.  \label{eqn:state evolution scalar form}
\end{align}

We emphasize that although the above results are obtained in the
asymptotic regimes where $N,K,L$ go to infinity, they can be used to
predict the performance of practical systems with finite but large
$N,K,L,M$ accurately.  In particular, for a practical system with
parameters $\rho^{\rm pilot}$, $L$, $K$, $N$ and pathloss $\beta_k$
for each user $k$, we simply set
\begin{align}
\label{eqn:parameters}
\xi  = L \rho^{\rm pilot}, \qquad
\epsilon = \frac{K}{N}, \qquad
\omega = \frac{N}{L},
\end{align}
in order to run the simplified state evolution
(\ref{eqn:state evolution scalar form 0})-(\ref{eqn:state evolution scalar form})
to obtain $\tau_\infty^2$ and subsequently $\upsilon_k$ and $\Delta \upsilon_k$ for each
user $k$. Although the above asymptotic results are obtained in the limit
of large $M$, they already corroborate well with the simulation results
as shown in Part I of this paper \cite{LiuPart1} for practical values of $M=16$ and $M=64$.
In this Part II of the paper, we assume the above characterization of
the channel estimation error in order to analytically characterize the
overall achievable rate.

\subsection{High SNR Characterization of Channel Estimation}

A key step in obtaining the statistics of the channel estimation error
according to (\ref{eqn:channel_estimate 1})--(\ref{eqn:channel_error 2})
is in identifying the fixed point $\tau_\infty^2$ of the state evolution
(\ref{eqn:state evolution scalar form}). In general, the fixed point
is a complicated function of the system parameter. But in certain
regime of practical interest, simple analytic characterization of the
fixed point can be obtained.

Observe that in practice, the vector AMP algorithm for device activity
detection and channel estimation should work in the regime of
$\omega\epsilon<1$, i.e., $L > K$, in order to control the channel
estimation error.  Thus, the behavior of $\tau_\infty^2$ when $L > K$
is of most interest. Further, the iterative state evolution simplifies
considerably in the high signal-to-noise ratio (SNR) limit.
The first technical result of this paper is a high SNR
characterization of the fixed point.

\begin{theorem}\label{corollary1}
Suppose that $\omega\epsilon<1$, i.e., $L>K$. Then, there is a unique fixed-point solution $\tau_\infty^2$ to (\ref{eqn:state evolution scalar form}), which satisfies
\begin{align}\label{eqn:bound on tau}
\frac{\sigma^2}{\xi}\leq \tau_\infty^2 \leq \frac{\sigma^2}{\xi (1-\omega \epsilon)}.
\end{align}
Moreover, suppose that the channel path-loss variable $\beta$ is bounded below, i.e.,
$\beta \ge \beta_{\rm min}$, for some positive $\beta_{\rm min}$.
Then, in the SNR regime where $\frac{\xi\beta_{\rm
min}}{\sigma^2}\rightarrow \infty$, the unique fixed-point solution to
(\ref{eqn:state evolution scalar form}) is given by
\begin{align}\label{eqn:tau infty}
\tau_\infty^2 \rightarrow \frac{\sigma^2}{\xi (1-\omega \epsilon)}.
\end{align}
\end{theorem}

\begin{IEEEproof}
Please refer to Appendix \ref{appendix8}.
\end{IEEEproof}

Note that in a single-cell system without inter-cell interference, the
SNR of the even cell-edge user is typically high within reasonable
range. As a result, the approximation of $\tau_\infty^2$ given in
(\ref{eqn:tau infty}) is expected to be accurate in the single-cell
system, as verified later in this paper by simulations. Moreover, (\ref{eqn:bound on tau}) shows that
the upper bound of $\tau_\infty^2$ is $\sigma^2/\xi (1-\omega \epsilon)$. Therefore, the asymptotic high-SNR limit obtained in (\ref{eqn:tau infty}) is also the worst-case noise power, and all the results based on this approximation can be viewed as performance lower bound for any value of SNR.

The main consequence of Theorem \ref{corollary1} is that under
practical system parameters $K$, $L$, $N$, $\rho^{\rm pilot}$, and for
reasonably large $M$ (such as $M=16$ or $64$), the covariance matrices of the estimated channel
and the channel estimation error for user $k$, resulting from the use of AMP
for joint device detection and channel estimation, are in the form of (\ref{eqn:channel_estimate 1}) and (\ref{eqn:channel_error 1}), in which $\upsilon_k$ and $\Delta \upsilon_k$ can be approximated
respectively as:
\begin{align}
\upsilon_k & =\frac{\beta_k^2}{\beta_k+
\frac{\sigma^2}{\rho^{\rm pilot} (L-K)}},
\label{eqn:channel estimate}
\end{align}
and
\begin{align}
\Delta \upsilon_k & =\frac{\beta_k
\frac{\sigma^2}{\rho^{\rm pilot} (L-K)}}{\beta_k+\frac{\sigma^2}{\rho^{\rm pilot} (L-K)}},
\label{eqn:channel estimation error}
\end{align}
where we have used (\ref{eqn:parameters}). Curiously, the above
expressions are independent of $N$. This is because device activity
detection is already perfect in the massive MIMO regime; the channel
estimation error is mainly due to the non-orthogonality of the pilot
sequences of the $K$ active users.

\section{Achievable Rate for Massive Connectivity}
\label{sec:Phase II: Data Transmission}

We are now ready to use 
the channel estimation error characterization in the previous section to
evaluate the achievable data transmission rate in the second
phase while accounting for the channel estimation error, in the
massive MIMO regime.  As user activity detection is perfect in
the massive MIMO regime in the first phase, we focus on an equivalent
wireless system in the second phase consisting of only $K$ active users
that simultaneously transmit their data to the BS in the uplink.
Moreover, for these users, we utilize the covariance matrices of
the estimated channels and channel estimation errors as given in
(\ref{eqn:channel_estimate 1})-(\ref{eqn:channel_error 2}), or
as in the high SNR regime,
(\ref{eqn:channel estimate})-(\ref{eqn:channel estimation error}).

\begin{figure*}[b]
\hrulefill
\setcounter{equation}{16}
\begin{equation}\label{eqn:SINR}
\gamma_k = \frac{\rho^{\rm data} |\mv{w}_k^H\hat{\mv{h}}_k|^2}{\rho^{\rm data} \sum\limits_{n\in \mathcal{K},n\neq k}|\mv{w}_k^H\hat{\mv{h}}_n|^2+\rho^{\rm data} \|\mv{w}_k\|^2\sum\limits_{n\in \mathcal{K}}\frac{\beta_n\tau_\infty^2}{\beta_n+\tau_\infty^2}+\sigma^2\|\mv{w}_k\|^2}.
\end{equation}
\setcounter{equation}{13}
\end{figure*}

In this paper, we choose to study the user achievable rate in certain
asymptotic regime, where not only $M$ goes to infinity, but also
$K$ goes to infinity, while their ratio is kept fixed, i.e.,
$K/M\rightarrow \mu$ with $\mu\in (0,\infty)$. Note that this is a
different asymptotic regime as in the analysis of the first phase, but
we justify by pointing out that both analyses are ultimately intended
for performance projection of system with finite parameters. Had we
followed the asymptotic regime of the analysis of the first phase,
where $K$ goes to infinity first for each finite $M$, then let $M$ go
to infinity, we would have obtained zero user rate,
which is not of practical interest.
Our present approach of letting both $K$ and $M$ go to infinity in
the analysis of the second phase, while simply assuming the channel
estimation characterization of the first phase, is validated by simulation
later in the paper. It also
leads to valuable system insight by allowing performance comparison
to the case with prior user activity information at the BS, i.e.,
the case with orthogonal pilot sequences assignment as widely assumed
in the current massive MIMO literature.

\subsection{Achievable Rates with MRC and MMSE Receivers}
\label{sec:achievable rate}

The equivalent baseband signal received at the BS for the second phase
is expressed as
\begin{align}\label{eqn:received signal phase 2}
\mv{y}=\sum\limits_{n\in \mathcal{K}}\mv{h}_n\sqrt{\rho^{\rm data} }s_n+\mv{z},
\end{align}
where $s_n\sim \mathcal{CN}(0,1)$ denotes the transmit symbol of user
$n\in \mathcal{K}$, which is modeled as a CSCG random variable with
zero-mean and unit-variance, $\rho^{\rm data}$ denotes the identical transmit
power of the active users in the second transmission phase, and
$\mv{z}\sim \mathcal{CN}(\mv{0},\sigma^2\mv{I})$ denotes the
AWGN at the BS.

The BS employs linear beamforming on the received signal $\mv{y}$ for decoding user messages:
\begin{align}\label{eqn:linear beamforming}
\hat{s}_k=&\mv{w}_k^H\left(\sum\limits_{n\in \mathcal{K}}\mv{h}_n\sqrt{\rho^{\rm data} }s_n+\mv{z}\right)\nonumber \\ =&\mv{w}_k^H\hat{\mv{h}}_k\sqrt{\rho^{\rm data} }s_k+\mv{w}_k^H\sum\limits_{n\in \mathcal{K},n\neq k}\hat{\mv{h}}_n\sqrt{\rho^{\rm data} }s_n\nonumber \\ &+\mv{w}_k^H\sum\limits_{n\in \mathcal{K}}\Delta \mv{h}_n\sqrt{\rho^{\rm data} }s_n+\mv{w}_k^H\mv{z}, ~~~ \forall k\in \mathcal{K},
\end{align}
where $\mv{w}_k\in \mathbb{C}^{M\times 1}$ denotes the beamforming vector for
the active user $k\in \mathcal{K}$.
In the above signal model, the BS views the estimated channels as the
true channels, and treats the term due to the channel estimation
error, i.e., $\mv{w}_k^H\sum_{n\in \mathcal{K}}\Delta
\mv{h}_n\sqrt{\rho^{\rm data} }s_n$, as additional noise.

Assume that the estimated channel and channel estimation error for
each active user $k$ are Gaussian distributed with the covariance matrices given in
(\ref{eqn:channel_estimate 1})-(\ref{eqn:channel_error 2}), i.e., $\hat{\mv{h}}_k\sim \mathcal{CN}(\mv{0},\frac{\beta_k^2}{\beta_k+\tau_\infty^2}\mv{I})$ and
$\Delta \mv{h}_k\sim \mathcal{CN}(\mv{0},\frac{\beta_k\tau_\infty^2}{\beta_k+\tau_\infty^2}\mv{I})$.
This can be justified by the fact that in the
asymptotic massive MIMO regime, user activity detection is perfect and
the MMSE denoiser as given in Theorem 1 of Part I asymptotically
becomes a linear MMSE channel estimator for the active users. As a
result, the estimated channels from the AMP algorithm can be assumed
to be close to Gaussian in the massive MIMO limit.
Following the standard bounding technique based on the worst case uncorrelated noise
\cite{hassibi_hochwald}, the uplink achievable rate of active user $k$
can be written down as
\begin{align}\label{eqn:achievable rate}
R_k=\frac{T-L}{T} \log_2(1+\gamma_k), ~~~ \forall k,
\end{align}where the signal-to-interference-plus-noise ratio (SINR) of
user $k$ given the channel realization is shown in (\ref{eqn:SINR}) on
the bottom of the page.

\setcounter{equation}{17}


This paper considers two different receive beamforming strategies,
namely the MRC beamforming and MMSE beamforming, which are
respectively defined as
\begin{align}
& \mv{w}_k^{{\rm MRC}}=\hat{\mv{h}}_k, \label{eqn:MRC} \\
& \mv{w}_k^{{\rm MMSE}}= \nonumber \\
& \quad \left(\sum\limits_{n\in \mathcal{K}}\rho^{\rm data} \hat{\mv{h}}_n\hat{\mv{h}}_n^H\hspace{-3pt}+\hspace{-3pt}\sum\limits_{n\in \mathcal{K}}\frac{\rho^{\rm data} \beta_n\tau_\infty^2}{\beta_n\hspace{-3pt}+\hspace{-3pt}\tau_\infty^2}\mv{I}\hspace{-3pt}+\hspace{-3pt}\sigma^2\mv{I}\right)^{-1}\hat{\mv{h}}_k. \label{eqn:MMSE}
\end{align}

The following theorem characterizes the achievable rates of each user
with the MRC beamforming and the MMSE beamforming, respectively, in our interested
asymptotic regime.

\begin{theorem}\label{theorem2}
Consider an uplink massive MIMO system with $M$ BS antennas serving $K$ users.
Assume that the estimated channel and channel estimation error for each active user
$k$ are Gaussian distributed with the covariance matrices given in
(\ref{eqn:channel_estimate 1})-(\ref{eqn:channel_error 2}), i.e., $\hat{\mv{h}}_k\sim \mathcal{CN}(\mv{0},\frac{\beta_k^2}{\beta_k+\tau_\infty^2}\mv{I})$ and
$\Delta \mv{h}_k\sim \mathcal{CN}(\mv{0},\frac{\beta_k\tau_\infty^2}{\beta_k+\tau_\infty^2}\mv{I})$, $\forall k\in \mathcal{K}$.
In the asymptotic regime where both $K,M$ go infinity but with their
ratio kept constant, i.e., $K/M\rightarrow \mu$ with $\mu \in (0,\infty)$,
the achievable rate for each user, assuming MRC beamforming (\ref{eqn:MRC}) at the BS,
is given by (\ref{eqn:achievable rate}), where
\begin{align}\label{eqn:SINR MRC}
\gamma_k^{\rm MRC} \rightarrow \frac{\beta_k^2}{\mu \mathbb{E}[\beta] (\beta_k+\tau_\infty^2)}, ~~~ \forall k.
\end{align}
%
%
The achievable rate for each active user,
assuming MMSE beamforming (\ref{eqn:MMSE}) at the BS, is given
by (\ref{eqn:achievable rate}), where
\begin{align}\label{eqn:SINR MMSE}
\gamma_k^{\rm MMSE} \rightarrow \frac{\beta_k^2}{\beta_k+\tau_\infty^2}
\Gamma, ~~~ \forall k,
\end{align}
with $\Gamma$ being the unique finite fixed-point solution of the following equation:
\begin{align}\label{eqn:Gamma}
\Gamma=\frac{1}{\mu\mathbb{E}\left[\frac{\beta^2}{\beta+\tau_\infty^2+\beta^2\Gamma}\right]
+\mu\mathbb{E}\left[\frac{\beta\tau_\infty^2}{\beta+\tau_\infty^2}\right]}.
\end{align}
\end{theorem}

\begin{IEEEproof}
Please refer to Appendix \ref{appendix3}.
\end{IEEEproof}

We remark that given the perfect user detection and channel estimation characterization obtained in Part I of this paper \cite{LiuPart1},
Theorem \ref{theorem2} can also be obtained based on the techniques used in \cite{debbah}. Since we are considering a single-cell rather than multi-cell setting,
we are able to provide a different and simpler proof in Appendix \ref{appendix3}. Note that this two-part paper studies a different system as compared to \cite{debbah}, since
in \cite{debbah} there are only $K$ users who are assumed to be always active, while in our paper, $K$ out of $N$ users are active in each coherence interval, as result, the device activity detection step has impact on the channel estimation error, thus leading to more involved SINR expressions as compare to \cite{debbah}.

We also remark that if the channel estimation had been perfect,
i.e., $\tau_\infty^2=0$ so that $\hat{\mv{h}}_k=\mv{h}_k$,
$\forall k$, the above theorem reduces to known results in the
literature. With the MRC receive beamforming at the BS, each user's
SINR given in (\ref{eqn:SINR MRC}) in this case reduces to
\begin{align}
\gamma_k^{\rm MRC} \rightarrow \frac{\beta_k}{\mu \mathbb{E}[\beta]}.
\end{align}
This is the same result as in \cite[Proposition 3.3]{tse_hanly}.

Moreover, with the MMSE receive beamforming at the BS, $\Gamma$
as given in (\ref{eqn:Gamma}) in the perfect channel estimation case
reduces to the fixed-point solution to the following equation:
\begin{align}\label{eqn:Gamma 1}
\Gamma=\frac{1}{\mu\mathbb{E}\left[\frac{\beta}{1+\beta\Gamma}\right]}.
\end{align}
As a result, each user's SINR is the fixed-point solution to the following equation:
\begin{align}
\gamma_k^{\rm MMSE}=\beta_k\Gamma  =
\frac{\beta_k}{\mu\mathbb{E}\left[\frac{\beta}{1+\beta\Gamma}\right]}
= \frac{\beta_k}{\mu\mathbb{E}\left[\frac{\beta\beta_k}{\beta_k+\beta\gamma_k^{\rm MMSE}}\right]},
\label{eqn:fixed-point SINR}
\end{align}
which is the same result as in \cite[Theorem 3.1]{tse_hanly}.

Observe that the user achievable rates under both the MRC
and MMSE beamforming strategies as shown in Theorem \ref{theorem2}
are finite, in contrast to the conventional single-cell
(thus without pilot contamination) massive MIMO scenario with a small
number of users, where the user achievable rates go to infinite in the
massive MIMO limit \cite{marzetta,larsson}.
This is because in a massive connectivity scenario where the number
of users is comparable with the number of antennas at the BS, the
total inter-user interference power seen by each user is comparable
to that of its desired signal, due to the fact that although each
interference alone is very weak due to the channel asymptotic
orthogonality, there are a large number of interference sources in the
system, resulting in finite achievable rate.


It is also worth noting that the MRC beamforming is optimal
in the conventional single-cell massive MIMO system in the asymptotic
limit of large number of BS antennas but finite number of
users, 
because the user channels become orthogonal with each other in the
limit, thus the inter-user interference is asymptotically zero. But
this is not the case for the massive connectivity scenario under
consideration in which the number of users also goes to infinity. Because
of the large number of interference sources in the system, the inter-user
interference remains significant with MRC beamforming. In
contrast, the MMSE beamforming strategy can more effectively control the
inter-user interference. As a result, there is a performance gap
between the MRC and MMSE beamforming strategies in the massive
connectivity scenario.

\subsection{User Scheduling for Overloaded System}
\label{sec:User Scheduling for Overloaded System}
\begin{figure}
\begin{center}
\scalebox{0.4}{\includegraphics*{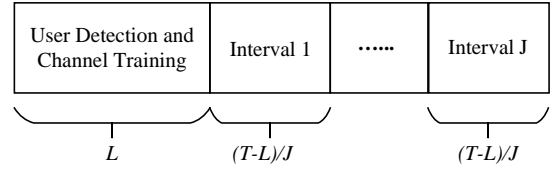}}
\end{center}
\caption{User scheduling strategy in an overloaded system.}\label{fig6}
\end{figure}

The above analysis assumes that in the second phase, all the $K$ active users
transmit simultaneously to the BS. It is worth noting that in an
overloaded system where the number of active users is larger than the
number of the antennas at the BS, i.e., $\mu=K/M>1$, in general we
should further divide the second phase into $J$ intervals such that in each
interval only $K/J$ users are scheduled for information transmission
in order to control the inter-user interference, as shown in
Fig.~\ref{fig6}. In the following, we formulate the user achievable
rates with scheduling in an overloaded system. The optimization over $J$ is treated later in Section \ref{sec:Optimization of User Scheduling}. Note that we assume a
finite $J$ such that $K/J$ goes to infinity thus Theorem
\ref{theorem2} still applies to each scheduled interval.

First, consider the case with the MRC beamforming at the BS.
Note that for each interval the ratio between the numbers of
the scheduled users and the antennas at the BS is reduced to
$K/(JM)=\mu/J$. Moreover, the transmission time\footnote{We ignore the
overhead for informing each active user of the index of its scheduled
interval since it is negligible compared to $L$.}
for each active user is reduced to $(T-L)/J$. As a result, the rate
expression for each active user becomes
\begin{align}\label{eqn:achievable rate TDMA}
R_k^{\rm MRC,SC}=\frac{T-L}{TJ}\log_2(1+\gamma_k^{\rm MRC,SC}), ~~~ \forall k,
\end{align}where the SINR is
\begin{align}\label{eqn:SINR TDMA}
\gamma_k^{\rm MRC,SC} \rightarrow \frac{J \beta_k^2}{\mu \mathbb{E}[\beta] (\beta_k+\tau_\infty^2)}.
\end{align}

Moreover, with the MMSE beamforming at the BS, the achievable rate for each active user is given by
\begin{align}\label{eqn:achievable rate MMSE TDMA}
R_k^{\rm MMSE,SC}=\frac{T-L}{TJ}\log_2(1+\gamma_k^{\rm MMSE,SC}), ~~~ \forall k,
\end{align}where the SINR is
\begin{align}\label{eqn:SINR MMSE TDMA}
\gamma_k^{\rm MMSE,SC} \rightarrow \frac{\beta_k^2}{\beta_k+\tau_\infty^2} \Gamma,
\end{align}with $\Gamma$ the fixed-point solution to
\begin{align}\label{eqn:SINR 1 MMSE TDMA}
\Gamma=\frac{J}{\mu\mathbb{E}\left[\frac{\beta^2}{\beta+\tau_\infty^2
+\beta^2\Gamma}\right]+\mu\mathbb{E}\left[\frac{\beta\tau_\infty^2}{\beta+\tau_\infty^2}\right]}.
\end{align}

\subsection{High SNR Approximation of User Rate}
\label{sec:Approximate User Rata}



When the overall system operates in the regime $L>K$, and if we assume
high SNR, we can use (\ref{eqn:tau infty}) to approximate
$\tau_\infty^2$ in the above rate expressions.
In this case, (\ref{eqn:tau infty}) can be further simplified
as $\tau_\infty^2=\frac{\sigma^2}{\rho^{\rm pilot}(L-K)}$. Further,
for practical systems with finite $K$ and $M$, expressions such as
$\mu\mathbb{E}[\beta]$ can be replaced by their emperical average,
i.e., $\frac{1}{M}\sum_{k\in \mathcal{K}}\beta_k$.


With the above approximations, for the case without user scheduling,
the user SINRs using the MRC and the MMSE receive
beamforming
as given in (\ref{eqn:SINR MRC}) and (\ref{eqn:SINR MMSE}), respectively,
reduce to
\begin{align}
& \gamma_k^{\rm MRC} \approx \frac{\beta_k^2}{\frac{1}{M}\sum\limits_{n\in \mathcal{K}}\beta_n (\beta_k+\frac{\sigma^2}{\rho^{\rm pilot} (L-K)})}, ~~~ \forall k, \label{eqn:SINR pilot MRC} \\
& \gamma_k^{\rm MMSE} \approx \frac{\beta_k^2}{\beta_k+\frac{\sigma^2}{\rho^{\rm pilot} (L-K)}}\Gamma, ~~~ \forall k, \label{eqn:SINR pilot MMSE}
\end{align}
with $\Gamma$ being the unique solution to the following equation:
\begin{multline}\label{eqn:SINR 1 pilot MMSE TDMA}
\frac{1}{\Gamma} =  \frac{1}{M}\sum\limits_{n\in \mathcal{K}}\frac{\beta_n^2}{\beta_n+\frac{\sigma^2}{\rho^{\rm pilot} (L-K)}
+\beta_n^2\Gamma} \\ +\frac{1}{M}\sum\limits_{n\in \mathcal{K}}\frac{\frac{\beta_n\sigma^2}{\rho^{\rm pilot} (L-K)}}{\beta_n+\frac{\sigma^2}{\rho^{\rm pilot} (L-K)}}.
\end{multline}
For the case with user scheduling, we have:
\begin{align}
& \gamma_k^{\rm MRC,SC} \approx \frac{J\beta_k^2}{\frac{1}{M}\sum\limits_{n\in \mathcal{K}}\beta_n (\beta_k+\frac{\sigma^2}{\rho^{\rm pilot} (L-K)})}, ~~~ \forall k, \label{eqn:SINR pilot MRC TDMA} \\
& \gamma_k^{\rm MMSE,SC} \approx \frac{\beta_k^2}{\beta_k+\frac{\sigma^2}{\rho^{\rm pilot} (L-K)}}\Gamma, ~~~ \forall k, \label{eqn:SINR pilot MMSE TDMA}
\end{align}
with $\Gamma$ being the unique solution to the following equation:
\begin{multline}\label{eqn:SINR 2 pilot MMSE TDMA}
\frac{J}{\Gamma}=\frac{1}{M}\sum\limits_{n\in \mathcal{K}}\frac{\beta_n^2}{\beta_n+\frac{\sigma^2}{\rho^{\rm pilot} (L-K)}
+\beta_n^2\Gamma} \\ +\frac{1}{M}\sum\limits_{n\in \mathcal{K}}\frac{\frac{\beta_n\sigma^2}{\rho^{\rm pilot} (L-K)}}{\beta_n+\frac{\sigma^2}{\rho^{\rm pilot} (L-K)}}.
\end{multline}
%

\section{Cost of Massive Device Detection }
\label{sec:The Cost of User Detection on Users' Rates}


One of the main results from Part I of this paper \cite{LiuPart1} is that in the
massive MIMO regime, user activity detection can always be made with
negligible probability of error. What is then the cost of device
detection? A goal of the Part II of this paper is to illustrate that
the cost of device detection arises as consequence of significantly
larger channel estimation error due to the use of non-orthogonal pilot
sequences.
This section quantifies such cost by comparing the user achievable rate
as given in the previous section 
to the achievable rate of the widely studied massive MIMO system
with known user activity but with imperfect channel estimation. We
focus on the $L>K$ regime in order to have reasonable channel
estimation error. For simplicity, we ignore the issue of scheduling
and assume that all active users transmit simultaneously in the second phase.


When the user activities are perfectly known at the BS, Phase I of
the transmission then consists of only the $K$ active users sending
their pilot sequences to the BS for channel estimation purpose.
Similar to (\ref{eqn:channel_training}), the received signal at the BS is
\begin{align}\label{eqn:received signal phase 1 known user activity}
\mv{Y}= & \sqrt{\rho^{\rm pilot} L}\sum\limits_{k\in \mathcal{K}}\mv{a}_k\mv{h}_k^H
+\mv{Z} \nonumber \\ =&\sqrt{\rho^{\rm pilot} L}\mv{A}_{\mathcal{K}}\mv{H}_{\mathcal{K}}+\mv{Z},
\end{align}where $\mv{A}_{\mathcal{K}}=[\cdots,\mv{a}_k,\cdots]\in \mathbb{C}^{L\times K}$
with $\|\mv{a}_k\|^2=1$ and $\mv{H}_{\mathcal{K}}=[\cdots,\mv{h}_k,\cdots]^H\in \mathbb{C}^{K\times M}$
are the collections of the pilot sequences and channels for all the active users $k\in \mathcal{K}$.

Differing, however, from the massive connectivity scenario 
where the pilot sequences must be non-orthogonal, e.g., the entries of
$\mv{A}$ in (\ref{eqn:channel_training}) are generated based on
the i.i.d.\ Gaussian distribution, in the case with prior user
activity information, it is the best to assign orthogonal pilot
sequences with length $L\geq K$ to the active users \cite{hassibi_hochwald},
i.e., $\mv{A}_{\mathcal{K}}^H\mv{A}_{\mathcal{K}}=\mv{I}$.
The BS then applies matching filter, i.e., $\mv{A}_{\mathcal{K}}^H$, to its
received signal (\ref{eqn:received signal phase 1 known user activity}),
resulting in
\begin{align}\label{eqn:indepedent channel known user activity}
\hat{\mv{h}}_k=\sqrt{\rho^{\rm pilot} L}\mv{h}_k+(\mv{a}_k^H\mv{Z})^H, ~~~ \forall k\in \mathcal{K}.
\end{align}
Note that the equivalent noise is distributed as $(\mv{a}_k^H\mv{Z})^H\sim \mathcal{CN}(\mv{0},\sigma^2\mv{I})$.
It can be shown that if the MMSE channel estimation is used on the channel
model (\ref{eqn:indepedent channel known user activity}), the estimated channels and
their uncorrelated channel estimation errors are distributed as
$\hat{\mv{h}}_k\sim
\mathcal{CN}\left(\mv{0},\frac{\beta_k^2}{\beta_k+\sigma^2/(\rho^{\rm
pilot}L)}\mv{I}\right)$
and $\Delta \mv{h}_k \sim
\mathcal{CN}\left(\mv{0},\frac{\beta_k\sigma^2/(\rho^{\rm
pilot}L)}{\beta_k+\sigma^2/(\rho^{\rm pilot} L)}\mv{I}\right)$,
$\forall k\in \mathcal{K}$, respectively \cite{larsson}. Similar to
Theorem \ref{theorem2} and by using the approximation
technique used in Section \ref{sec:Approximate User Rata},
the users' rates achieved by the MRC and MMSE beamforming strategies
in the regime $L>K$ can be shown to be as given in (\ref{eqn:achievable rate}), where
\begin{align}
& \gamma_k^{\rm MRC}\approx \frac{\beta_k^2}{\frac{1}{M}\sum\limits_{n\in \mathcal{K}}\beta_n (\beta_k+\frac{\sigma^2}{\rho^{\rm pilot} L})}, ~~~ \forall k, \label{eqn:SINR known user activity MRC} \\
& \gamma_k^{\rm MMSE} \approx \frac{\beta_k^2}
{\beta_k+\frac{\sigma^2}{\rho^{\rm pilot} L}} \Gamma, ~~~ \forall k, \label{eqn:SINR known user activity MMSE}
\end{align}
with $\Gamma$ being the unique solution to the following equation:
\begin{multline}
\frac{1}{\Gamma}=\frac{1}{M}\sum\limits_{n\in \mathcal{K}}\frac{\beta_n^2}{\beta_n+\frac{\sigma^2}{\rho^{\rm pilot} L}
+\beta_n^2\Gamma} \\ +\frac{1}{M}\sum\limits_{n\in \mathcal{K}} \frac{\frac{\beta_n\sigma^2}{\rho^{\rm pilot} L}}{\beta_n+\frac{\sigma^2}{\rho^{\rm pilot} L}}.
\end{multline}

Comparing to the massive connectivity scenario without prior user activity
information, for which the SINRs achieved by the MRC and MMSE beamforming
are given in (\ref{eqn:SINR pilot MRC}) and (\ref{eqn:SINR pilot MMSE}),
respectively, 
it can be observed that the cost of user activity detection lies in
the effective channel estimation error, which increases from
$\frac{\sigma^2}{\rho^{\rm pilot} L}$ to $\frac{\sigma^2}{\rho^{\rm pilot} (L-K)}$.

As mentioned earlier, the reason for this cost is that for the
massive connectivity scenario, since $L<N$, it is impossible to assign
orthogonal pilot sequences to all $N$ users. If the entries of
$\mv{A}$ are generated according to i.i.d.\ Gaussian distribution,
although the user activity detection by the vector AMP algorithm is
perfect due to the results in Part I of this paper \cite{LiuPart1}, this choice of $\mv{A}$
nevertheless results in larger channel estimation error because of
multiuser interference as compared to the case where orthogonal
pilot sequences can be used. This is reminiscent of the well-known
inter-cell pilot contamination problem in conventional massive MIMO
systems, except that the contamination now comes from the non-orthongal pilots
within the cell as the cost of supporting massive connectivity.


\section{Optimization of Pilot Length}
\label{sec:The Optimal Length of Pilot Sequences}

The characterization of the channel estimation error and user
achievable rates also allows an optimization of the pilot sequence
length for maximizing the system sum rate. 
Longer pilot sequences result in better channel estimation but
shorter data transmission time, and vice versa, so there is an optimal
$L$ that balances the two effects. 
Again in this section, we ignore scheduling and assume that all
active users transmit simultaneously in the second phase.
The optimization of user scheduling is discussed in the next section.

First, consider the case with MRC beamforming at the BS. According to
(\ref{eqn:achievable rate}) and
(\ref{eqn:SINR pilot MRC}), in the practical regime of $L>K$, the sum rate maximization problem can be expressed as
\begin{align}
\mathop{\mathrm{max}}_{K < L < T} & ~ \frac{T-L}{T}\sum\limits_{k\in \mathcal{K}}
	\log_2\left(1+\frac{M\beta_k^2}{\sum\limits_{n\in \mathcal{K}}\beta_n
	(\beta_k+\frac{\sigma^2}{\rho^{\rm pilot} (L-K)})}\right)
\label{eqn:problem P2}
\end{align}

\begin{theorem}\label{theorem5}
The objective function of problem (\ref{eqn:problem P2}) is a
concave function over $L$ in the range $K<L<T$, if $L$ is relaxed as
a real number.
\end{theorem}

\begin{IEEEproof}
Please refer to Appendix \ref{appendix5}.
\end{IEEEproof}

According to Theorem \ref{theorem5}, problem (\ref{eqn:problem P2})
can be globally solved as follows. First, we ignore the constraint
that $L$ is an integer and solve the relaxed convex version of problem
(\ref{eqn:problem P2}). Let $L^\ast$ denote the optimal solution,
which is not necessarily an integer. Then, $L^\ast$ either rounding up
or rounding down to the next integer value would be the optimal pilot
sequence length, depending on which way maximizes the user sum rate.

Next, consider the case when the MMSE beamforming is employed at the BS.
According to (\ref{eqn:achievable rate}) and (\ref{eqn:SINR pilot MMSE}), in the case of $L>K$, the sum rate maximization problem over the
pilot sequence length for the MMSE beamforming case is
\begin{align}
\mathop{\mathrm{max}}_{K < L < T} & ~ \frac{T-L}{T} \sum\limits_{k\in \mathcal{K}}
\log_2\left(1+\frac{\beta_k^2}{\beta_k+\frac{\sigma^2}{\rho^{\rm pilot}(L-K)}}
	\Gamma\right) 
\label{eqn:problem P3}
\end{align}
where $\Gamma$ is the solution to (\ref{eqn:SINR 1 pilot MMSE TDMA}).
However, since $\Gamma$ is a complicated function of $L$, it is
non-trivial to solve the problem (\ref{eqn:problem P3}).
Nevertheless, the optimal pilot sequence length for the MMSE
beamforming case can be obtained by a one-dimension search.

\section{Optimization of User Scheduling}
\label{sec:Optimization of User Scheduling}

We now consider the question of in an overloaded system with more
users than the number of BS antennas, what the optimal number of
scheduling intervals $J$ should be chosen as for maximizing the systme
sum rate.
Assuming $L>K$, consider first the case of MRC beamforming at the BS.
According to the user rate expressions given in
(\ref{eqn:achievable rate TDMA}) and (\ref{eqn:SINR pilot MRC TDMA}),
the sum rate maximization problem over $J$ can be formulated as
\begin{align}
\mathop{\mathrm{max}}_{J \ge 1} & ~ \frac{T-L}{TJ} \sum\limits_{k\in \mathcal{K}} \log_2\left(1+\frac{JM \beta_k^2}{\sum\limits_{n\in \mathcal{K}}\beta_n (\beta_k+\frac{\sigma^2}{\rho^{\rm pilot} (L-K)})}\right)
\label{eqn:problem P4}
\end{align}

\begin{theorem}\label{theorem4}
The objective function of problem (\ref{eqn:problem P4}) is a monotonically decreasing function over $J$. As a result, the optimal solution
to problem (\ref{eqn:problem P4}) is $J^\ast=1$.
\end{theorem}

\begin{IEEEproof}
Please refer to Appendix \ref{appendix1}.
\end{IEEEproof}

Intuitively, Theorem \ref{theorem4} implies that under MRC, if we
reduce the number of scheduled users in each interval, the sacrifice
of data transmission time plays a more significant role on user sum
rate than the reduction in inter-user interference. Such a phenomenon
reveals the inefficiency of MRC beamforming in an overloaded system,
since even user scheduling cannot improve the user sum rate.

Next, consider the case when the MMSE beamforming is employed at the BS.
According to the user rate given in (\ref{eqn:achievable rate MMSE TDMA})
and (\ref{eqn:SINR pilot MMSE TDMA}), the sum rate maximization
problem over $J$ can be formulated as
\begin{align}\mathop{\mathrm{max}}_{J\ge 1} & ~ \frac{T-L}{TJ} \sum\limits_{k\in \mathcal{K}} \log_2\left(1+\frac{\beta_k^2}{\beta_k+\frac{\sigma^2}{\rho^{\rm pilot} (L-K)}} \Gamma\right) \label{eqn:problem P5}
\end{align}
where $\Gamma$ is the solution to (\ref{eqn:SINR 2 pilot MMSE TDMA}). Since the solution to
(\ref{eqn:SINR 2 pilot MMSE TDMA}) is a complicated function of $J$,
it is non-trivial to solve problem (\ref{eqn:problem P5}) analytically.
However, the optimal solution to problem (\ref{eqn:problem P5}) can be
easily obtained numerically via a one-dimension search.

Differing from the case of MRC beamforming at the BS, as
shown later by numerical simulations,
the optimal solution to problem (\ref{eqn:problem P5}) is $J$ strictly
larger than $1$ in general. Thus, user scheduling can significantly
improve the user sum rate when the MMSE beamforming is employed at the BS.

\section{Numerical Examples}\label{sec:Numerical Examples}

In this section, we provide numerical examples to verify the main
results of this paper. The setup is the same as in the numerical
simulations in Part I of this paper. There are $N=2000$ users in a
single cell.  Let $d_n$ denote the distance between user $n$ and the BS,
$\forall n$. It is assumed that $d_n$'s are randomly distributed in
the regime $[0.05{\rm km},1{\rm km}]$.  The path loss model of the
wireless channel for user $n$ is given as
$\beta_n=-128.1-36.7\log_{10}(d_n)$ in dB, $\forall n$.  The bandwidth
and the coherence time of the wireless channel are $1$MHz and $1$ms,
respectively, thus in each coherence block $T=1000$ symbols can be
transmitted. The transmit power for each user at both the first and
second transmission phases is $\rho^{\rm pilot}=\rho^{\rm data}
=23$dBm. The power spectral density of the AWGN at the BS is assumed
to be $-169$dBm/Hz. Moreover, all the following numerical results are
obtained by averaging over $100,000$ channel realizations.

\subsection{Fixed-Point of State Evolution for AMP}

\begin{figure}
\begin{center}
\scalebox{0.6}{\includegraphics*{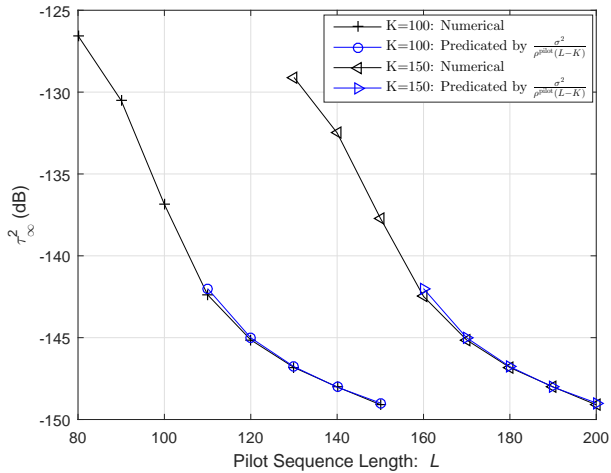}}
\end{center}
\caption{Fixed point of state evolution when each of the $N=2000$ users accesses the channel with probability $\epsilon=0.05$ or $\epsilon=0.075$ in each coherence time; the BS has $M=128$ antennas, and the SNR of the farthest user is $14$dB.}\label{fig9}
\end{figure}

Fig.~\ref{fig9} shows the numerical evaluation of the fixed-point
solution to the state evolution of AMP, which is used for
characterizing the channel estimation error.
In this numerical example, each of the $N$ users accesses the channel with probability $\epsilon=0.05$ or $\epsilon=0.075$ in each
coherence time (around $K=100$ or $K=150$ users are active), and the number of antennas at the BS is $M=128$. Note
that in this example, the SNR of the farthest user, which is $1$km
away from the BS, is $14$dB. Fig.~\ref{fig9} shows the comparison
between the numerical evaluation of the fixed point
(\ref{eqn:state evolution scalar form}) and the high-SNR approximation
given in (\ref{eqn:tau infty}) in Theorem \ref{corollary1}
for different values of $L$. Note that the transmit power is set to be
$\xi=L\rho^{\rm pilot}$ so that (\ref{eqn:tau infty}) reduces to
$\tau_\infty^2=\frac{\sigma^2}{\rho^{\rm pilot}(L-K)}$. It is observed
that (\ref{eqn:tau infty}) is a very good approximation of the exact
fixed-point solution in this practical SNR range when $L>K$.

\subsection{Cost for User Activity Detection on User Rates}
\begin{figure}
\begin{center}
\subfigure[MRC Receive Beamforming with $M=128$]{\scalebox{0.6}{\includegraphics*{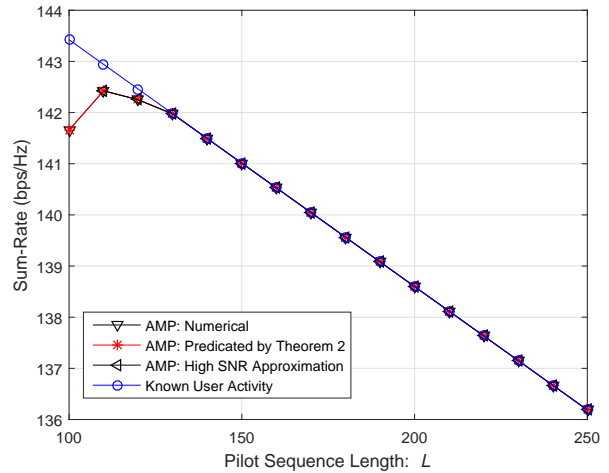}}} \\
\subfigure[MRC Receive Beamforming with $M=256$]{\scalebox{0.6}{\includegraphics*{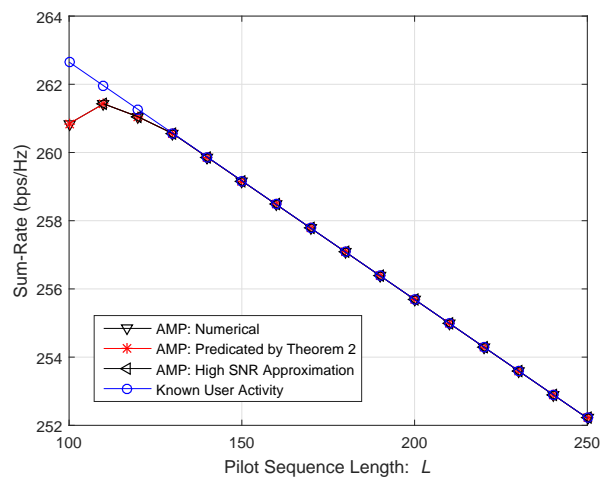}}} \\
\caption{User sum rate comparison with MRC receive beamforming between
the cases with and without prior user activity information
at the BS when each of the $N=2000$ users
accesses the channel with probability $\epsilon=0.05$ in each coherence time and the BS has $M=128$ or $256$ antennas.}\label{fig4}
\end{center}
\end{figure}

\begin{figure}
\begin{center}
\subfigure[MMSE Receive Beamforming with $M=128$]{\scalebox{0.6}{\includegraphics*{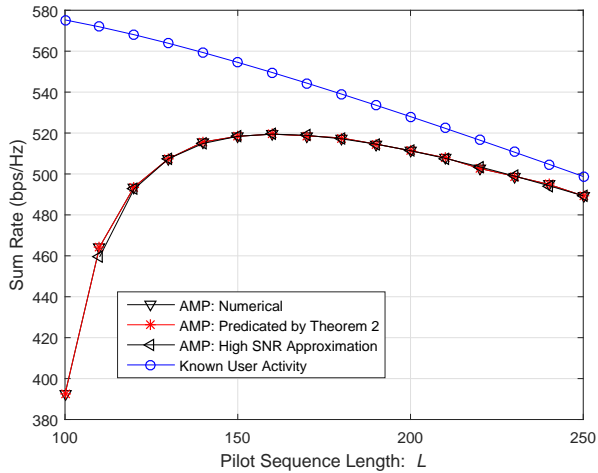}}} \\
\subfigure[MMSE Receive Beamforming with $M=256$]{\scalebox{0.6}{\includegraphics*{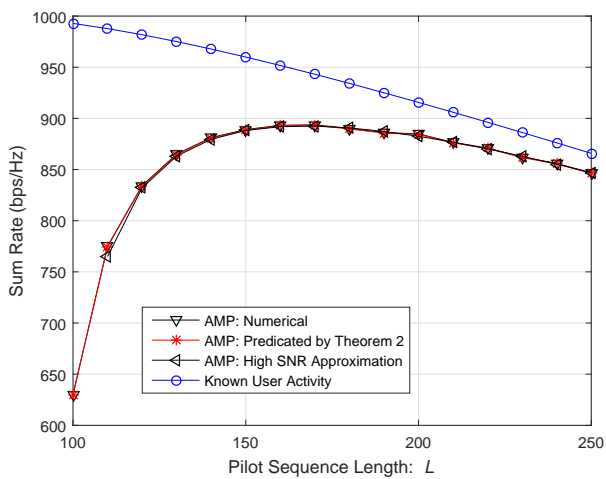}}} \\
\caption{User sum rate comparison with MMSE receive beamforming
between the cases with and without prior user activity information
at the BS when each of the $N=2000$ users
accesses the channel with probability $\epsilon=0.05$ in each
coherence time and the BS has $M=128$ or $M=256$
antennas.}\label{fig_known_activity}
\end{center}
\end{figure}

Next we quantify the cost of user activity detection on achievable
rates. Figs.~\ref{fig4} and \ref{fig_known_activity} show the user sum rates versus
the length of the pilot sequences $L$ for both the cases of MRC and
MMSE beamforming at the BS. In this numerical example, there are
$M=128$ or $M=256$ antennas at the BS and each of the $N=2000$ users accesses the channel
with probability $\epsilon=0.05$ at
each coherence time (around $K=100$ users are active). As baseline, the scenario with prior information
on user activity known at the BS is also plotted, where orthogonal
pilot sequences can be assigned to the active users for channel
estimation in the first phase.

With the MRC beamforming at the BS, it is observed from
Fig.~\ref{fig4} that for the case without prior information
of the user activity, the theoretical result shown
in Theorem \ref{theorem2} and the high-SNR approximation
(\ref{eqn:achievable rate}) and (\ref{eqn:SINR pilot MRC})
both perfectly match the numerical result for various values of $L$.
Moreover, it is observed that the optimal pilot lengths are
$L=K=100$ and $L=110$ for the cases with and without prior information
of the user activity at the BS, respectively. Note that without prior
information of the user activity, the MSE for channel estimation is
larger, thus more time needs to be spent in the first phase to improve
the channel estimation accuracy. Finally, it is observed that maximal
sum rates for the cases with and without prior information of user
activity at the BS are very close, indicating that the cost of user
activity detection is quite small under MRC beamforming.


The user achievable rate can be dramatically improved, however,
if MRC beamforming is replaced with MMSE beamforming, as shown in
Fig.~\ref{fig_known_activity}.
It can be observed from Fig.~\ref{fig_known_activity} that with the
MMSE beamforming at the BS, the theoretical result shown in Theorem
\ref{theorem2} and the high-SNR approximation (\ref{eqn:achievable rate})
and (\ref{eqn:SINR pilot MMSE})-(\ref{eqn:SINR 1 pilot MMSE TDMA})
perfectly match the numerical result for all values of $L$. Moreover,
it is observed that the optimal pilot length is $L=160$ when user
activity is not known a priori at the BS, and the cost of user
activity detection is about 10\% of the overall sum rate. Note that
this optimal length is much longer than that for the case with MRC
beamforming, which is $L=110$. This is because different from the MRC
beamforming, the MMSE beamforming for each user is a function of the
estimated channels of all the users, as shown in (\ref{eqn:MMSE}). As
a result, the performance of MMSE beamforming is more sensitive to the
channel estimation error, thus we should allocate more time for
channel training.  Given the significant sum rate improvement of MMSE
beamforming over MRC beamforming, this is a small price to pay.

It is worth emphasizing that MRC is not well suited for massive
connectivity applications, because as explained earlier it is unable
to mitigate the significant multiuser interference stemmed from a
large number of devices. The fact that MMSE beamforming is capable of
achieving five or six times higher sum rate than MRC, as shown in
Figs.~\ref{fig4} and \ref{fig_known_activity}, illustrates that MMSE
rather than MRC beamforming should be used for massive
connectivity applications, even though MRC would have been adequate
in conventional massive MIMO systems.



\subsection{The Impact of User Scheduling on User Rates}\label{sec:The Impact of User Scheduling on Users' Rates}
\begin{figure}
\begin{center}
\scalebox{0.6}{\includegraphics*{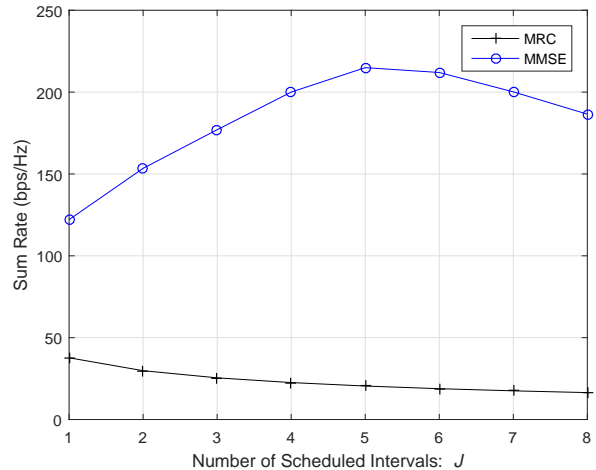}}
\end{center}
\caption{User sum rate versus different numbers of scheduled intervals in the second
transmission phase when each of the $N=2000$ users
accesses the channel with probability $\epsilon=0.15$ in each coherence
time and the BS has $M=64$ antennas.}\label{fig5}
\end{figure}

Finally, we study the impact of user scheduling in an
overloaded system. In this example, we assume that there are $M=64$
antennas at the BS and each of the $N=2000$ users
accesses the channel with probability $\epsilon=0.15$ at each coherence
time (around $K=300$ users are active) such that $\mu=\frac{K}{M}>1$. It is further assumed that the
pilot sequence length is $L=400$. Fig.~\ref{fig5} shows the
user sum rate versus the number of scheduled intervals of the
data transmission phase $J$. For the case of MRC beamforming
at the BS, it is observed that the user sum rate decreases with
the number of scheduled intervals $J$, which verifies Theorem
\ref{theorem4}. This is in fact an indication of the inefficiency of
MRC beamforming in an overloaded system. In contrast, for the
MMSE beamforming, it is observed that user scheduling can
significantly enhance the overall sum rate. Specifically, in this
numerical example the optimal strategy is to schedule $60$ 
users in each of $J=5$ intervals such that for any particular interval
the system is almost fully loaded. This example shows that for massive
connectivity applications with massive MIMO, if the number of users is
much larger than the number of antennas at the BS, combining user
scheduling together with MMSE receive beamforming at the BS can be a
good strategy for managing multiuser interference.

\section{Conclusion and Future Work}\label{sec:Conclusion}

This two-part paper illustrates that massive MIMO is ideally suited
for massive connectivity applications. The main technical contribution
of the overall two-part paper is a characterization of the effect of
using non-orthogonal pilot sequences for massive device activity
detection, channel estimation, and data transmission. The main
conclusion of this Part II of the paper is that despite perfect device
activity detection in the massive MIMO regime,
a loss in the overall achievable transmission rate nevertheless
arises as compared to the conventional massive MIMO system because
of the significantly larger channel estimation error due to the
non-orthogonality of pilot sequences. We also show that for massive
connectivity applications, it is essential to use MMSE beamforming
instead of MRC; the optimal pilot length should be longer than that
in conventional massive MIMO systems in order to compensate for the
additional channel estimation error; finally scheduling can enhance
the overall transmission rate.

There are a number of directions along which the results of this paper
can be further extended. First, we mention that power control has not
been taken into account. In this paper, all the active users transmit
with an identical transmit power in each of the first and second phases.
It is
conceivable that users far away from the BS can be assigned with higher
power so that a more fair rate distribution among all the active users
can be achieved. Second,
this paper has not addressed the issue of optimal scheduling. Future work on how to select the active
users in each scheduled interval to maximize user achievable rate will be of interest.
Further, the results of this paper are restricted to single-cell
scenarios.
Future work can extend the existing results to account for inter-cell
interference and to investigate ways to provide adequate coverage
to cell-edge users.

\begin{appendix}
\subsection{Proof of Theorem \ref{corollary1}}\label{appendix8}

First, we show that when $\omega \epsilon < 1$, the fixed point of the
simplified state evolution
\begin{align}\label{eqn:state evolution 2}
\tau_{\infty}^2=\frac{\sigma^2}{\xi } +
\omega \epsilon
\mathbb{E}_{\beta}\left[\frac{\beta\tau_\infty^2}{\beta+\tau_\infty^2}\right],
\end{align}
is unique. Define
\begin{align}\label{eqn:fx}
f(x)=x-\omega\epsilon\mathbb{E}_\beta\left[\frac{\beta
x}{\beta+x}\right]-\frac{\sigma^2}{\xi}, ~~~ x\geq 0.
\end{align}
It can be easily shown that $f(x)$ is a continuous function of $x$.
Moreover, the derivative of $f(x)$ is
\begin{align}
f'(x)=1-\omega\epsilon\mathbb{E}_\beta\left[\frac{\beta^2}{(\beta+x)^2}\right],
~~~ x\geq 0.
\end{align}
When $\omega\epsilon< 1$, we have $f'(x)\geq 0$,
thus $f(x)$ is a monotonically increasing function for $x\in [0,\infty)$.
Consequently, the fixed point of (\ref{eqn:state evolution 2}) is unique.

Second, we show that the fixed-point solution of (\ref{eqn:state evolution 2}) is bounded by (\ref{eqn:bound on tau}) if $\omega\epsilon< 1$. It can be easily seen from (\ref{eqn:state evolution 2})
that $\tau_\infty^2\geq \sigma^2/\xi$. Moreover, we have
\begin{align}
\tau_{\infty}^2=\frac{\sigma^2}{\xi } +
\omega \epsilon
\mathbb{E}_{\beta}\left[\frac{\beta\tau_\infty^2}{\beta+\tau_\infty^2}\right]\leq \frac{\sigma^2}{\xi } + \omega \epsilon \tau_\infty^2,
\end{align}where the inequality is because $\beta\tau_\infty^2/(\beta+\tau_\infty^2)\leq \tau_\infty^2$, $\forall \beta\geq 0$. As a result, if $\omega\epsilon< 1$, it follows that $\tau_\infty^2\leq \sigma^2/((1-\omega \epsilon)\xi)$.

Next, we verify that when
$\frac{\xi\beta_{\rm min}}{\sigma^2}\rightarrow \infty$,
\begin{align} \label{eqn:tau infty 2}
\tau_\infty^2 = \frac{\sigma^2}{\xi (1-\omega \epsilon)}
\end{align}
is a fixed-point solution of (\ref{eqn:state evolution 2}).
Substituting the above $\tau_\infty^2$ into the right-hand side of the
simplified state evolution, we have
\begin{align}
\frac{\sigma^2}{\xi } +
\omega \epsilon
\mathbb{E}_{\beta}\left[\frac{\beta\tau_\infty^2}{\beta+\tau_\infty^2}\right]
& = \frac{\sigma^2}{\xi } +
 \frac{\sigma^2\omega \epsilon}{\xi(1-\omega \epsilon) } \mathbb{E}_{\beta}\left[\frac{1}{1+\frac{\sigma^2}{\beta\xi(1-\omega \epsilon) }}\right]
\nonumber \\ & \rightarrow   \frac{\sigma^2}{\xi } +
\frac{\sigma^2\omega \epsilon}{\xi(1-\omega \epsilon) } \nonumber \\ & =
\frac{\sigma^2}{\xi(1-\omega \epsilon) } = \tau_\infty^2,
\end{align}
where the second last line is due to the high SNR assumption
and that $\omega \epsilon < 1$.
This verifies that (\ref{eqn:tau infty 2}) is the unique solution to
(\ref{eqn:state evolution 2}) in the high SNR limit.


\subsection{Proof of Theorem \ref{theorem2}}\label{appendix3}

With the MRC beamforming given in (\ref{eqn:MRC}), it can be shown that the SINR of user $k$ given in (\ref{eqn:SINR}) reduces to
\begin{align}\label{eqn:SINR MRC 1}
& \noindent \gamma_k^{\rm MRC} = \nonumber \\ & \frac{\rho^{\rm data} \|\hat{\mv{h}}_k\|^4}{\rho^{\rm data} \hspace{-2pt}\sum\limits_{n\in \mathcal{K},n\neq k}|\hat{\mv{h}}_k^H\hat{\mv{h}}_n|^2\hspace{-2pt}+\hspace{-2pt}\rho^{\rm data} \|\hat{\mv{h}}_k\|^2\hspace{-2pt}\sum\limits_{n\in \mathcal{K}}\frac{\beta_n\tau_\infty^2}{\beta_n+\tau_\infty^2}\hspace{-2pt}+\hspace{-2pt}\sigma^2\|\hat{\mv{h}}_k\|^2}.
\end{align}
If the estimated channels are distributed as $\hat{\mv{h}}_n\sim \mathcal{CN}(\mv{0},\frac{\beta_n^2}{\beta_n+\tau_\infty^2}\mv{I})$, $\forall n\in \mathcal{K}$, as $M\rightarrow \infty$, it thus follows that
\begin{align}
& \frac{\|\hat{\mv{h}}_k\|^2}{M}\rightarrow \left(\frac{\beta_k^2}{\beta_k+\tau_\infty^2}\right), \label{eqn:signal power}
\end{align}
and
\begin{align}
& \frac{\|\hat{\mv{h}}_k\|^2\sum\limits_{n\in \mathcal{K}}\frac{\beta_n\tau_\infty^2}{\beta_n+\tau_\infty^2}}{KM} \rightarrow \mathbb{E}\left[ \frac{\beta\tau_\infty^2}{\beta+\tau_\infty^2}\right]\left(\frac{\beta_k^2}{\beta_k+\tau_\infty^2}\right).
\end{align}
Moreover, according to Appendix B in \cite{tse_hanly}, we have
\begin{align}\label{eqn:interference}
\frac{\sum\limits_{n\in \mathcal{K},n\neq k}|\hat{\mv{h}}_k^H\hat{\mv{h}}_n|^2}{KM} \rightarrow \mathbb{E}\left[\frac{\beta^2}{\beta+\tau_\infty^2}\right]\frac{\beta_k^2}{\beta_k+\tau_\infty^2}.
\end{align}As a result, as $M\rightarrow \infty$, each active user's achievable SINR converges to
\begin{align}
& \noindent \gamma_k^{\rm MRC} \nonumber \\ & \rightarrow \frac{\rho^{\rm data} M^2\left(\frac{\beta_k^2}{\beta_k+\tau_\infty^2}\right)^2}{\rho^{\rm data} KM\mathbb{E}\left[\frac{\beta^2}{\beta+\tau_\infty^2}\hspace{-3pt}+\hspace{-3pt}\frac{\beta\tau_\infty^2}{\beta+\tau_\infty^2}\right]\left(\frac{\beta_k^2}{\beta_k+\tau_\infty^2}\right)\hspace{-3pt}+\hspace{-3pt}\sigma^2M\left(\frac{\beta_k^2}{\beta_k+\tau_\infty^2}\right)} \nonumber \\ & \rightarrow \frac{\beta_k^2}{\mu \mathbb{E}[\beta] (\beta_k+\tau_\infty^2)}, ~~~ \forall k,
\end{align}
thus establishing (\ref{eqn:SINR MRC}).

With the MMSE beamforming given in (\ref{eqn:MMSE}), it can be shown
that the SINR of user $k$ given in (\ref{eqn:SINR}) reduces to

\begin{multline}\label{eqn:SINR MMSE 1}
\gamma_k^{\rm MMSE}=\rho^{\rm data} \hat{\mv{h}}_k^H
\left(\sum\limits_{n\in \mathcal{K},n\neq k}\rho^{\rm data} \hat{\mv{h}}_n\hat{\mv{h}}_n^H \right. \\
\left. +\sum\limits_{n\in \mathcal{K}}\frac{\rho^{\rm data} \beta_n\tau_\infty^2}{\beta_n+\tau_\infty^2}\mv{I}+\sigma^2\mv{I}\right)^{-1}\hat{\mv{h}}_k.
\end{multline}


If the estimated channels are distributed as $\hat{\mv{h}}_k\sim \mathcal{CN}(\mv{0},\frac{\beta_k^2}{\beta_k+\tau_\infty^2}\mv{I})$, $\forall k\in \mathcal{K}$, as $M\rightarrow \infty$, it thus follows that
\begin{align}
& \noindent \gamma_k^{\rm MMSE} \nonumber \\ & \rightarrow \frac{\rho^{\rm data}
\beta_k^2}{M(\beta_k+\tau_\infty^2)}{\rm
tr}\bigg(\bigg(\sum\limits_{n\in \mathcal{K},n\neq k}\frac{\rho^{\rm
data} \hat{\mv{h}}_n\hat{\mv{h}}_n^H}{M} \nonumber \\ & \ \ \ \ \ \ \
\ \ \ \ \ \ \ \ \ \ \ \ \ +\sum\limits_{n\in
\mathcal{K}}\frac{\rho^{\rm data}
\beta_n\tau_\infty^2}{M(\beta_n+\tau_\infty^2)}\mv{I}+\frac{\sigma^2}{M}\mv{I}\bigg)^{-1}\bigg)
\label{eqn1} \\ & \rightarrow \frac{\rho^{\rm data}
\beta_k^2}{M(\beta_k+\tau_\infty^2)}{\rm
tr}\bigg(\bigg(\sum\limits_{n\in \mathcal{K}}\frac{\rho^{\rm data}
\hat{\mv{h}}_n\hat{\mv{h}}_n^H}{M} \nonumber \\ & \ \ \ \ \ \ \ \ \ \
\ \ \ \ \ \ \ \ \ \ +\sum\limits_{n\in \mathcal{K}}\frac{\rho^{\rm
data}
\beta_n\tau_\infty^2}{M(\beta_n+\tau_\infty^2)}\mv{I}+\frac{\sigma^2}{M}\mv{I}\bigg)^{-1}\bigg)
\label{eqn2}
\end{align}
\begin{align}
& \rightarrow \frac{\rho^{\rm data}
\beta_k^2}{M(\beta_k+\tau_\infty^2)}{\rm
tr}\bigg(\bigg(\sum\limits_{n\in \mathcal{K}}\frac{\rho^{\rm
data}\beta_n^2 }{M(1+e_n)(\beta_n+\tau_\infty^2)}\mv{I} \nonumber \\ &
\ \ \ \ \ \ \ \ \ \ \ \ \ \ \ \ \ \ \ \ +\sum\limits_{n\in
\mathcal{K}}\frac{\rho^{\rm data}
\beta_n\tau_\infty^2}{M(\beta_n+\tau_\infty^2)}\mv{I}+\frac{\sigma^2}{M}\mv{I}\bigg)^{-1}\bigg)
\label{eqn3} \\ & \rightarrow
\frac{\beta_k^2}{\beta_k+\tau_\infty^2}\cdot \frac{1}{\mu
\mathbb{E}\left[\frac{\beta^2}{(1+e)(\beta+\tau_\infty^2)}\right]+\mu
\mathbb{E}\left[\frac{\beta\tau_\infty^2}{\beta+\tau_\infty^2}\right]
+\frac{\sigma^2}{M\rho^{\rm data} }} \\ & \rightarrow
\frac{\beta_k^2}{\beta_k+\tau_\infty^2}\cdot \frac{1}{\mu \mathbb{E}\left[\frac{\beta^2}{(1+e)(\beta+\tau_\infty^2)}\right]+\mu \mathbb{E}\left[\frac{\beta\tau_\infty^2}{\beta+\tau_\infty^2}\right]},
\end{align}
where
\begin{align}
e_k=& \frac{1}{M}{\rm tr}\bigg(\mathbb{E}\left(\rho^{\rm data}\hat{\mv{h}}_k\hat{\mv{h}}_k^H\right)\bigg(\sum\limits_{n\in \mathcal{K}}\frac{\mathbb{E}\left(\rho^{\rm data}\hat{\mv{h}}_n\hat{\mv{h}}_n^H\right)}{M(1+e_n)} \nonumber \\ & \ \ \ \ \ \ \ +\sum\limits_{n\in \mathcal{K}}\frac{\rho^{\rm data} \beta_n\tau_\infty^2}{M(\beta_n+\tau_\infty^2)}\mv{I}+\frac{\sigma^2}{M}\mv{I}\bigg)^{-1}\bigg) \nonumber \\ \rightarrow& \frac{\rho^{\rm data} \beta_k^2}{M(\beta_k+\tau_\infty^2)}{\rm tr}\bigg(\bigg(\sum\limits_{n\in \mathcal{K}}\frac{\rho^{\rm data}\beta_n^2 }{M(1+e_n)(\beta_n+\tau_\infty^2)}\mv{I} \nonumber \\ & \ \ \ \ \ \ \ \ \ \ \ \ \ \ \ \ \ \ +\sum\limits_{n\in \mathcal{K}}\frac{\rho^{\rm data} \beta_n\tau_\infty^2}{M(\beta_n+\tau_\infty^2)}\mv{I}+\frac{\sigma^2}{M}\mv{I}\bigg)^{-1}\bigg) \nonumber \\ \rightarrow & \gamma_k^{\rm MMSE}. \label{eqn4}
\end{align}
In the above, (\ref{eqn1}) is due to \cite[Lemma 4]{debbah12}, (\ref{eqn2}) is due to \cite[Lemma 6]{debbah12}, (\ref{eqn3}) is due to \cite[Theorem 1]{debbah12}, and (\ref{eqn4}) is due to (\ref{eqn3}).

As a result, the user SINRs are the fixed-point solution to the following equations:
\begin{align}\label{eqn:SINR equations}
\gamma_k^{\rm MMSE} = & \frac{\beta_k^2}{\beta_k+\tau_\infty^2}\cdot \nonumber \\ & \frac{1}{\mu \mathbb{E}\left[\frac{\beta^2}{(1+\gamma^{\rm MMSE})(\beta+\tau_\infty^2)}+\frac{\beta\tau_\infty^2}{\beta+\tau_\infty^2}\right]}, ~~~ \forall k\in \mathcal{K}.
\end{align}
Define
\begin{align}
\Gamma=\frac{1}{\mu \mathbb{E}\left[\frac{\beta^2}{(1+\gamma^{\rm MMSE})(\beta+\tau_\infty^2)}+\frac{\beta\tau_\infty^2}{\beta+\tau_\infty^2}\right]}.
\end{align}Then, (\ref{eqn:SINR equations}) reduces to
\begin{align}\label{eqn:SINR equations 1}
\gamma_k^{\rm MMSE}=\frac{\beta_k^2}{\beta_k+\tau_\infty^2}\cdot \Gamma, ~~~ \forall k\in \mathcal{K}.
\end{align}

By taking (\ref{eqn:SINR equations 1}) into both the left-hand side and right-hand side of the equation given in (\ref{eqn:SINR equations}), it can be shown that $\Gamma$ is the fixed-point solution to the following equation:
\begin{align}
\Gamma=&\frac{1}{\mu \mathbb{E}\left[\frac{\beta^2}{(1+\frac{\beta^2}{\beta+\tau_\infty^2}\times \Gamma)(\beta+\tau_\infty^2)}+\frac{\beta\tau_\infty^2}{\beta+\tau_\infty^2}\right]} \\ = & \frac{1}{\mu\mathbb{E}\left[\frac{\beta^2}{\beta+\tau_\infty^2+\beta^2\Gamma}\right]+\mu\mathbb{E}\left[\frac{\beta\tau_\infty^2}{\beta+\tau_\infty^2}\right]}.
\end{align}

At last, we prove the uniqueness of the fixed-point solution to (\ref{eqn:Gamma}). First, it can be observed
that $\Gamma=0$ is not the fixed-point solution. Divide both the left-hand side and right-hand side
of (\ref{eqn:Gamma}) by $\Gamma$ and consider the following function:
\begin{align}
f(\Gamma)=\frac{1}{\mu\mathbb{E}\left[\frac{\beta^2\Gamma}{\beta+\tau_\infty^2+\beta^2\Gamma}\right]+\mu\mathbb{E}\left[\frac{\beta\tau_\infty^2\Gamma}{\beta+\tau_\infty^2}\right]}-1.
\end{align}
It can be shown that $f(\Gamma)$ is a decreasing function over $\Gamma$ when $\Gamma\geq 0$. Moreover, we have $f(\Gamma \rightarrow \infty)\rightarrow -1<0$ and $f(\Gamma=0)\rightarrow \infty>0$. As a result, the must be a unique finite solution to $f(\Gamma)=0$, which is the unique finite fixed-point solution to (\ref{eqn:Gamma}).

Theorem \ref{theorem2} is thus proved.

\subsection{Proof of Theorem \ref{theorem5}}\label{appendix5}
Suppose $L$ is relaxed as a real number. For convenience, define
\begin{align}
& f_k(L)=\log_2\left(1+\frac{\beta_k^2}{\frac{1}{M}\sum\limits_{n\in \mathcal{K}}\beta_n (\beta_k+\frac{\sigma^2}{\rho^{\rm pilot} (L-K)})}\right), \\
& f(L)=\sum\limits_{k\in \mathcal{K}} f_k(L), \\
& g(L)=\frac{T-L}{T}f(L).
\end{align}Note that $g(L)$ is the objective function of problem (\ref{eqn:problem P2}).

First, we study the function $f_k(L)$. Define
\begin{align}
a_k=\beta_k^2+\frac{1}{M}\sum\limits_{n\in \mathcal{K}}\beta_n\beta_k, ~~~ \forall k\in \mathcal{K}.
\end{align}
It can be shown that the first-order derivative of $f_k(L)$ is
\begin{align}
& \noindent f_k'(L) \nonumber \\
& = \frac{\frac{\beta_k^2\sigma^2}{\rho^{\rm pilot} \log 2}}{\left(a_k(L-K)+\frac{\frac{1}{M}\sum\limits_{n\in \mathcal{K}}\beta_n\sigma^2}{\rho^{\rm pilot} }\right)\left(\beta_k(L-K)+\frac{\sigma^2}{\rho^{\rm pilot} }\right)} \nonumber \\ & > 0, ~~~ {\rm if} ~ L>K.
\end{align}Moreover, it can be observed that $f_k'(L)$ is a monotonically decreasing function of $L$ if $L>K$. As a result, it follows that $f_k''(L)<0$, $\forall k$. It then follows that $f'(L)>0$ and $f''(L)<0$ when $L>K$.

Next, we study the function $g(L)$. It can be shown that the first and second-order derivatives of $g(L)$ are
\begin{align}
& g'(L)=\frac{-f(L)+(T-L)f'(L)}{T}, \\
& g''(L)=\frac{-2f'(L)+(T-L)f''(L)}{T}.
\end{align}Since $f'(L)>0$ and $f''(L)<0$, it then follows that $g''(L)<0$ when $L/K>1$. As a result, if $L$ is relaxed as a real number, the objective function of problem (\ref{eqn:problem P2}) is a concave function of $L$ when $L/K>1$. Theorem \ref{theorem5} is thus proved.

\subsection{Proof of Theorem \ref{theorem4}}\label{appendix1}
For convenience, define $x=1/J$. Then, according to (\ref{eqn:achievable rate TDMA}) and (\ref{eqn:SINR TDMA}), the rate of user $k$ is given as
\begin{align}
f_k(x)=\frac{(T-L)x}{T}\log_2\left(1+\frac{M\beta_k^2}{\sum\limits_{n\in \mathcal{K}}\beta_n (\beta_k+\frac{\sigma^2}{\rho^{\rm pilot}(L-K)}) x}\right).
\end{align}
Let us first ignore the constraint that $J$ is an integer, thus $x$ is a continuous variable. In this case, it can be shown that the first-order derivative of $f_k(x)$ is
\begin{align}
f_k'(x)=& \frac{T-L}{T}\log_2\left(1+\frac{\beta_k^2}{\frac{1}{M}\sum\limits_{n\in \mathcal{K}}\beta_n (\beta_k+\frac{\sigma^2}{\rho^{\rm pilot}(L-K)}) x}\right) \nonumber \\ & -\frac{T-L}{T\ln2}\frac{\beta_k^2}{\frac{1}{M}\sum\limits_{n\in \mathcal{K}}\beta_n(\beta_k+\frac{\sigma^2}{\rho^{\rm pilot}(L-K)})x+\beta_k^2}.
\end{align}Moreover, the second-order derivative of $f_k(x)$ is
\begin{align}
f_k''(x)&= -\frac{T-L}{T\ln 2}\frac{\beta_k^4}{[\frac{1}{M}\sum\limits_{n\in \mathcal{K}}\beta_n(\beta_k+\frac{\sigma^2}{\rho^{\rm pilot}(L-K)})x+\beta_k^2]^2x} \nonumber \\ & <0.
\end{align}As a result, $f_k'(x)$ is a decreasing function of $x$. It
can be shown that $f_k'(x\rightarrow \infty)\rightarrow 0$. It then
follows that $f_k'(x)>f_k'(x \rightarrow \infty)=0$, i.e., $f_k(x)$ is
an increasing function of $x$. Note that $x=1/J$, it thus follows that
each user's rate is a decreasing function of $J$. 
In other words, $J=1$ maximizes each user's rate. Consequently, the objective function of problem (\ref{eqn:problem P4}) is a decreasing function over $J$ and the optimal solution to problem (\ref{eqn:problem P4}) is $J^\ast=1$. Theorem \ref{theorem4} is thus proved.

\end{appendix}

\end{document}